# Effect of nonadiabatic spin transfer torque on domain wall resonance frequency and mass

Mahdi Jamali,[1] Kyung-Jin Lee,[2] and Hyunsoo Yang[1,a]

[1]Department of Electrical and Computer Engineering, National University of Singapore, 4 Engineering Drive 3, Singapore 117576, Singapore

[2]Department of Materials Science and Engineering, Korea University, Seoul 136-713, Republic of Korea

The dynamics of a magnetic domain wall in a semi circular nanowire loop is studied by an analytical model and micromagnetic simulations. We find a damped sinusoidal oscillation of the domain wall for small displacement angles around its equilibrium position under an external magnetic field in the absence of currents. By studying the effect of current induced nonadiabatic spin transfer torque on the magnetic domain wall resonance frequency and mass, a red shift is found in the resonance frequency and domain wall mass increases by increasing the ratio of nonadiabatic spin torque to adiabatic contribution above 1.

[a] e-mail address: eleyang@nus.edu.sg

Current induced motion of a magnetic domain wall (DW) in a ferromagnetic nanowire has been intensively studied recently due to its potential applications for next generation memories[1-3]. The determination of the DW mass is fundamentally interesting and also technologically important, since it limits the ultimate operational speed of the devices. The DW mass, first introduced by Döring[4], has been reported in the studies of current induced DW motion as well as in the presence of magnetic field[5-10]. Although there is a general consensus that the DW mass is directly correlated with the tilting angle and non-zero tilting angle plays the role of momentum[11], the experimental determination for the DW mass is less clear inferring the DW mass from $5.6 \times 10^{-25}$ to $6.55 \times 10^{-23}$ kg. One reason of this uncertainty is the tilting angle is directly affected by the nonadiabaticity of the current induced spin transfer torque[11], and the magnitude of the ratio ($\beta/\alpha$) of nonadiabatic spin torque $\beta$ to adiabatic one is still highly controversial which is distributed widely from 0 to 8, where $\alpha$ is the Gilbert damping constant[12-18]. In this Letter, we report analytical and micromagnetic studies of a free damped oscillation of magnetic DWs in a semi ring nanowire under an applied magnetic field without any current excitation which can affect the tilting and the mass of DWs. In this study the DW mass is deduced from the resonance frequency. In the case of sinusoidal current excitation with the field, the DW resonance frequency and mass is significantly affected by the current driven spin torque nonadiabaticity.

The structure that we have used in the simulations is shown in Fig. 1(a). The wire has a width of 100 nm, thickness of 10 nm, and the radius of curvature is 450 nm. The simulation cell size is $4 \times 4 \times 10$ nm$^3$ and the nanowire is made of Permalloy (Py) with the saturation magnetization ($M_S$) of $860 \times 10^3$ A/m, the exchange stiffness ($A_{ex}$) of $1.3 \times 10^{-11}$ J/m, and $\alpha$ of 0.01. We used the object oriented micromagnetic framework (OOMMF) code for simulations which incorporates the spin transfer torque term for the DW motion.[19] By applying a magnetic field of $H_y = 10$ kOe in the y-direction and relaxing the system, a single tail to tail transverse DW is generated in the center of the nanowire in Fig. 1(a). By applying a magnetic field of 200 Oe in the x-direction ($H_x$), the DW is displaced to the left side of the loop shown in Fig. 1(b). When the system is released in the presence of $H_y$, the DW undergoes a damped oscillatory motion similar to a pendulum in a gravity field.[6] A typical example is shown in Fig. 1(d) and (e) with $H_y$ = 200 Oe.

The magnetic transverse wall (TW) motion can be described based on the DW position parameter ($q$) and DW tilting angle ($\phi$) towards out of the plane. In a semi circular structure, $q = r\theta$ where $r$ is the circle radius and $\theta$ is the displacement angle from the equilibrium position in Fig. 1(c). Furthermore, one can decompose the applied magnetic field ($H_a$) to the tangential component ($H_t$) and the radial component ($H_r$). In the case of small $\theta$ and $\phi$, only the $H_t$ component can displace transverse DWs and it can be written as $H_t = H_0 \sin(\theta) = H_0 \theta$ in which $H_0$ is the amplitude of the applied magnetic field ($|H_a| = H_0$). One can write the transverse wall equations of motion based on the displacement angle parameter as follows:

$$\frac{r(1+\alpha^2)}{\Delta}\ddot{\theta} + \gamma_0 \alpha \left(H_0 + H_k \frac{r}{\Delta}\right)\dot{\theta} + \gamma_0^2 H_k H_0 \theta = \frac{\dot{u}}{\Delta}(1+\alpha\beta) + \gamma_0 H_k \frac{\beta u}{\Delta} \qquad (1)$$

where $\gamma_0$ is the gyromagnetic ratio, $H_k$ is the transverse anisotropy, $\Delta$ is the DW width, and the $u$ parameter is the effective drift velocity of the conduction electron spins defined by $u = JPg\mu_B/(2eM_S)$, where $J$ is the current density, $P$ is the spin polarization, $\mu_B$ is the Bohr magneton, and $e$ is the electron charge[20-22].

If the current is zero, the equation of motion is the form of a damped harmonic oscillator, $\ddot{\theta} + 2\rho\omega_0 \dot{\theta} + \omega_0^2 \theta = 0$ and the solution can be written as:

$$\theta(t) = \theta_0 \exp(-\rho\omega_0 t)\sin(\omega_0\sqrt{1-\rho^2}\,t + \Psi)$$
(2)

$$\omega_0^2 = \frac{\gamma_0^2 H_k H_0 \Delta}{(1+\alpha^2)r}, \quad \rho = \frac{\alpha(H_k + H_0\frac{\Delta}{r})}{2\sqrt{H_k H_0 \frac{\Delta}{r}(1+\alpha^2)}}$$

where $\omega_0$ is the undamped resonance frequency, $\rho$ is the damping ratio, and $\theta_0$ and $\psi$ are decided from the initial conditions.

Both $\theta$ and $\phi$ show a damped oscillatory behavior in Fig. 1(e) and (f) for small $\theta$ ($< \pm 6°$). The oscillatory motion of $\theta$ is fitted by a damped sinusoidal form described above as shown in Fig. 1(e) and it matches well with the data for small $\theta$. When a DW reaches to the peaks of $\theta$, the DW velocity is zero and $\phi$ is almost zero. On the other hand, when the DW is passing through its equilibrium position ($\theta = 0$), the DW has its maximum velocity and $\phi$ shows peak values as shown in Fig. 1(f). For large $\theta$, DW motion is very complicated. The Walker breakdown can be the reason for the complicated motion of a DW for large $\theta$.

Figure 2(a) shows the displacement profile of a DW for three different fields of $H_y = 150$, 200, and 300 Oe. By increasing $H_y$, the time for a DW to reach the zero angle position decreases due to a higher restoring force. The oscillation frequencies are $f = 1.8$, 1.91, and 2.08 GHz and the settling times, in which $\theta$ changes from 6 to 0.6°, are 3.93, 4.32, and 4.53 ns, for 150, 200, and 300 Oe, respectively. In order to calculate the mass ($m$) of a DW, one can use the formula of $f^2 = Q_m H/(4\pi^2 mr)$,[6, 10] where $f$ is the resonance frequency of a DW, $Q_m = 2\mu_0 M_s S$ is magnetic charge, $M_s$ is the saturation magnetization, $S$ is the cross sectional area of the nanowire, $H$ is the applied magnetic field in the y-direction, and $r$ is the radius of the loop. The resonance frequency in Eq. (2) can be linked to the DW mass by: $f = \frac{1}{2\pi}\sqrt{\frac{K}{m}}$, where $K = 2\mu_0 M_s H_0 S / r$ and

$$m = 2(1+\alpha^2)\mu_0 M_s S /(\gamma_0^2 H_k \Delta)$$
.

For $H_y$ less than 300 Oe, $f^2$ versus the magnetic field curve shows a linear relationship, however, for a large magnetic field, it deviates from the linear curve possibly due to the large disturbance of the nanowire potential profile[5, 6, 23]. With $S = 1\times10^{-15}$ m$^2$ and $r = 450$ nm, the DW mass is $1.27\times10^{-24}$ kg, which is in line with other reports[6, 8, 10]. In order to obtain the DW mass for a vortex wall, a semi ring with the width of 250 nm, thickness of 30 nm, and radius of curvature 625 nm was simulated. For $H_y = 200$ Oe, a resonance frequency of 352 MHz was obtained resulting in the mass of $8.43\times10^{-23}$ kg.

Since the DW width is proportional to $\sqrt{A_{ex}/K_a}$, where $K_a$ is the magnitude of anisotropy constant due to shape in our case, and the DW mass is inversely proportional to the DW width, the resonance frequency is proportional to $\sqrt[4]{A_{ex}}$. Figure 2(c) shows the effect of the exchange stiffness of ferromagnetic materials on the resonance frequency of DWs. It is clear that by increasing the exchange stiffness, the frequency of oscillation increases and the DW mass reduces. By increasing the exchange stiffness, the overshoot of the oscillation increases and the settling times for $A_{ex} = 1.0\times10^{-11}$, $1.4\times10^{-11}$, and $2.0\times10^{-11}$ J/m are 3.16, 3.87, and 4.25 ns, respectively. Figure 2(d) shows the DW displacement angle profile for the three different values of radius of curvature; $r = 450$, 550, and 650 nm. The resonance frequency and overshoot of the displacement angle reduce as the loop radius increases which is in line with Eq. (2).

In order to study the effect of nonadiabatic coefficient on the DW resonance frequency and mass, we have injected the sinusoidal current into the nanowire and measured its resonance frequency. From Eq. (1) the current induced spin torque transfer cannot change the resonance frequency of the system, which shows one of the limitations in the one dimensional (1D) model. The structure for our simulation is shown in the inset of Fig. 2(e). It is a part of semi circular nanowire with a width of 100 nm and thickness of 10 nm. We have injected the sinusoidal current density of $2\times10^8$ A/cm$^2$ and sweep the frequency. We also applied a magnetic field of 200 Oe in the y-direction during simulations. As can be seen in Fig. 2(e), there is no change in the resonance frequency as long as $\beta/\alpha \leq 1$, whereas for $\beta/\alpha >1$ there is a red shift in the DW resonance frequency. A red shift in the frequency leads to an increase in the DW mass. The same trend has been observed in the maximum DW tilting angle in Fig. 2(f). By increasing $\beta/\alpha$, the DW tilting angle increases and DW momentum is enhanced. In the case of $\beta/\alpha \leq 1$, the resonance frequency of nanowire was measured to be 1.55 GHz (with both the current and field excitation), while the resonance frequency that measured from a damped oscillation in Fig. 1 is 1.914 GHz for a magnetic field of 200 Oe (only with field). This frequency shift leads to a sizable change (52%) in the DW mass. The shift of the resonance frequency in a high current density regime cannot be explained based on the 1D model in which the resonance frequency is independent of the current density. The micromagnetic simulations provide more details DW dynamics in this regime than the 1D model.

We replace the half of a semi loop with a straight nanowire as shown in Fig. 3(a). The width (100 nm) and thickness (10 nm) of the straight wire is the same as the curved part and the length of the straight wire is 2 μm. For the creation of a DW, a magnetic field of 10 kOe at ~45° to the straight part of the nanowire is applied and then the system is relaxed. With $H_y = 200$ Oe, the DW accelerates and the DW moves into the straight part of the nanowire due to its finite momentum. We remove the external field as soon as the DW enters into the straight part (at 1.3 ns). As time progresses, the displacement of a DW in the straight wire decreases since a drag force on the DW dissipates the kinetic energy of the DW. The finite displacement of a DW in the straight wire can be explained by the finite mass and momentum of a DW. The DW position versus time curve in Fig. 3(b) is fitted by an exponential curve $a\exp(-t/\tau)+b$ derived from 1D model with the parameters of $a = -4.9 \times 10^{-6}$ m, $\tau = 1.8$ ns, and $b = 6.7 \times 10^{-7}$ m. From the relationship, the time constant $\tau = (1+\alpha^2)/(\gamma_0 H_k \alpha)$, the transverse anisotropy $H_k$ is estimated to be 711 Oe.

In conclusion, we study the DW motion in a semi loop of ferromagnetic nanowire theoretically and numerically. A pendulum like motion and finite penetration of the DW into the straight wire is attributed to the finite mass of the magnetic DW and the mass is determined without involving the complexity of the nonadiabaticity. The study of the effect of nonadiabatic coefficient on the DW resonance shows that there is a red shift in the DW resonance frequency increasing the DW mass, if the current excitation is used with the field. When the ratio of $\beta/\alpha$ increases more than 1, the DW mass also increases inferred from the red shift in the resonance frequency. Our results show that the nonadiabaticity of the current driven spin transfer torque can affect the DW resonance frequency and mass significantly and this observation cannot be explained within the simple 1D model. The shift of the resonance frequency in the presence of currents can be utilized to determine the magnitude of the nonadiabatic coefficient, which is under considerable debate.

This work is supported by the Singapore National Research Foundation under CRP Award No. NRF-CRP 4-2008-06.

Figure Captions

FIG. 1. (a) A DW generated by applying $H_y=10$ kOe and relaxing the system. (b) Displacing the DW from the initial position by applying $H_x$. (c) Displacement angle ($\theta$) and decomposition of the applied field. A DW releasing with $H_y$ is equivalent to that of a pendulum in a gravity field. (d) Time evolution of the DW dynamics for $H_y=200$ Oe. (e) A damped sinusoidal curve fitting of $\theta$ for $H_y=200$ Oe. (f) Domian wall displacement ($\theta$) and tilting angle ($\phi$) profile for $H_y=200$ Oe.

FIG. 2. (a) $\theta$ profile for three different magnetic fields. (b) Oscillation frequency ($f$) of a DW in the different magneitc fields. (c) DW oscillation frequency and mass for different exchange stiffnesses. (d) $\theta$ profile for the different radii. (e) The maximun $\theta$ versus ac current frequency. (f) The maximun $\phi$ versus ac current frequency.

FIG. 3. (a) Time evolution of the DW dynamics in a rounded L-shape structure. (b) The DW position versus time curve with an exponential fit. The origin of the displacement is at 1.3 ns when the external field is removed.

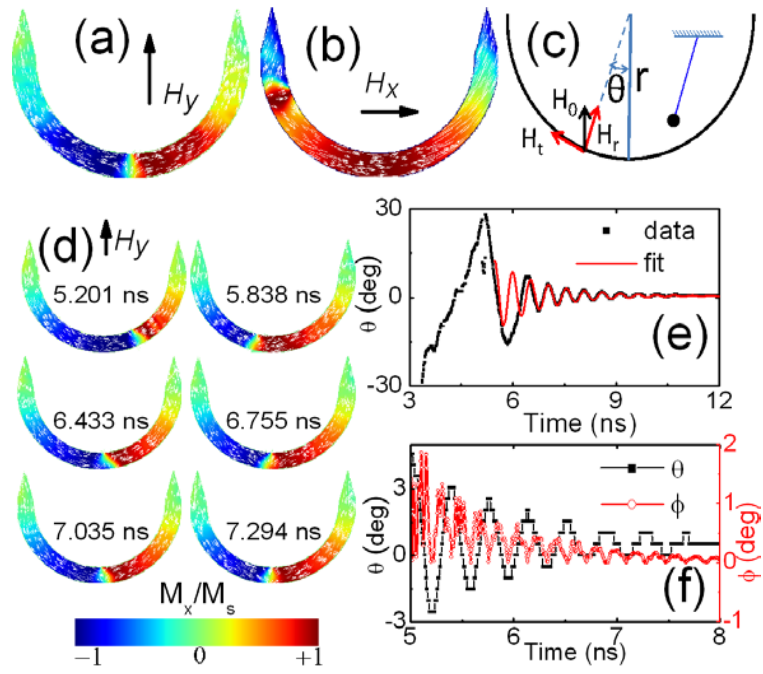

Figure 1.

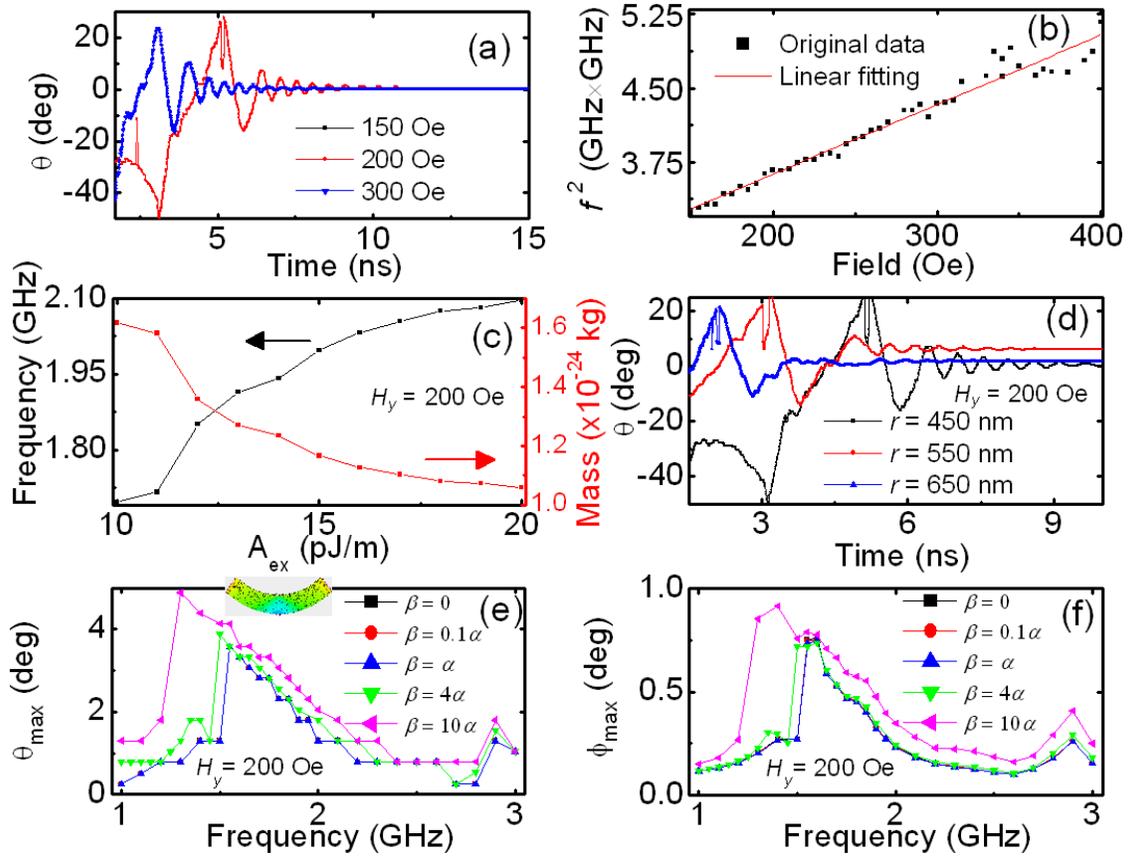

Figure 2.

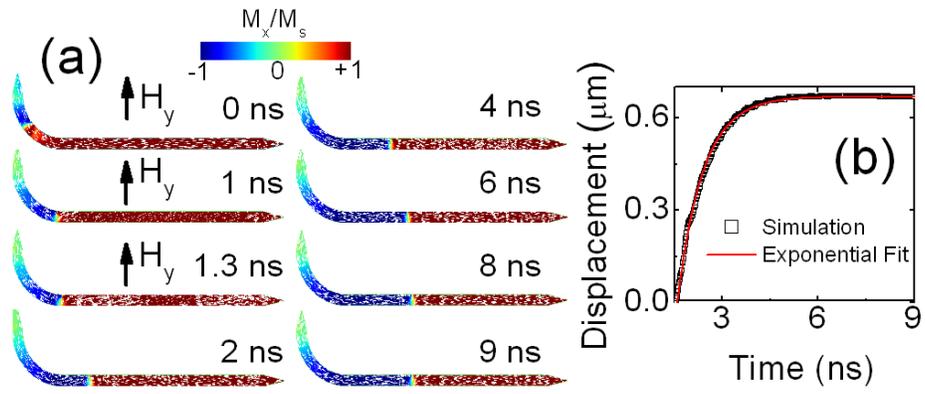

Figure 3.